\documentclass[twocolumn,showpacs,floatfix,aps]{revtex4}
\usepackage{graphicx}
\usepackage{dcolumn}
\usepackage{bm}

\begin{document}

\title{Entanglement versus chaos in disordered spin chains}

\author{L. F. Santos}
\email{santos@pa.msu.edu} 
\affiliation{Department of Physics and Astronomy, Michigan State
  University, East Lansing, MI 48824}
\author{G. Rigolin}
\email{rigolin@ifi.unicamp.br}
\author{C. O. Escobar}
\email{escobar@ifi.unicamp.br}
\affiliation{Departamento de Raios C\'osmicos e Cronologia, Instituto de 
F\'{\i}sica Gleb Wataghin, Universidade Estadual de Campinas, C.P. 6165, 
cep 13084-971, Campinas, S\~ao Paulo, Brazil}

\begin{abstract}
We use a Heisenberg spin-1/2 chain to 
investigate how chaos and localization 
may affect the entanglement of pairs of qubits. To 
measure how much entangled a pair is, we compute its concurrence, which is
then analyzed in the delocalized/localized and in the chaotic/non-chaotic regimes. Our results indicate that
chaos reduces entanglement and that
entanglement decreases in the region of strong localization.
In the transition region from a chaotic to a non-chaotic regime localization increases entanglement. We also show that entanglement is larger for strongly interacting qubits (nearest neighbors) than for weakly interacting qubits (next and next-next neighbors).  
\end{abstract}


\maketitle

\newpage

\section{Introduction}

Today it is well known that quantum entanglement, hereafter simply entanglement, is not just an issue of quantum mechanics that has yielded several discussions on the conceptual foundations of this theory \cite{epr, schroedinger, bell}. It is also a practical resource to develop new technologies. Entanglement enables us to implement some quantum algorithms which outperform their classical counterparts \cite{shor, grover}. It can also increase the amount of classical information transmitted in a quantum channel via the super dense coding protocol \cite{superbennett} and it allows the transmission of an unknown quantum state between two spatially separated parties (quantum teleportation) \cite{telebennett, zeilinger}. 

Since we can perform many useful tasks with entangled states, it is desirable to quantify the amount of entanglement these states have. There are at least three measures of entanglement which have reasonable physical interpretations \cite{formation, destilation, relative}. One of them, the Entanglement of Formation ($E_{F}$), has a relatively simple analytical expression for bipartite mixed states \cite{wootters}. The $E_{F}$ is a monotonically increasing function of a quantity called concurrence, which we adopt here as our measure of entanglement \cite{horodecki}. See section II for details.        

In this paper, we study how the entanglement 
of pairs of qubits is related to chaos and localization.
To do so, we consider an one 
dimensional Heisenberg spin-1/2 chain. The interest in such systems has recently 
increased considerably, for they model several proposed quantum computers \cite{mark}. Each spin in the chain corresponds to a qubit and the 
interaction between them is used to describe two-qubit gates.
A single-particle excitation corresponds to a spin pointing up. In these systems, we call defect the site whose energy splitting differs from all the others. It is obtained by applying a different magnetic field in the $z$ direction to the chosen site. A disordered system is characterized by the presence of one or more defects.

There have been recent attempts to relate localization
with entanglement \cite{hu} and also chaos with entanglement \cite{bet_shepe,fazio}.
Here we study at the same time the influence of localization and chaos 
on the entanglement. Some authors have shown that in a system of coupled 
quantum kicked tops \cite{Miller,Bandyopadhyay} chaos increases entanglement.
Here we show just the opposite, that chaos {\it decreases} entanglement.
Even though our system is very different from the one they considered,
it is clear that a more careful analysis of the relation
between chaos and entanglement is still needed. We claim that 
the entanglement
of two qubits is not simply related to delocalization. It also is 
influenced by the chaoticity of the system. By chaos we mean how close 
the energy level spacing distribution of the system is to a 
Wigner-Dyson distribution. See next paragraph and Section III for details.

The spin chain considered in this paper is a perfect system to 
study the relations between entanglement and 
localization and between entanglement and chaos, 
because it has different regions of interest. 
Concurrences of pairs of qubits obtained in different regimes can then
be compared.
The transition from integrability to
non-integrability in these systems depends on the defects \cite{integ_lea}.
It is known that the energy level spacing distribution of an integrable 
system is Poissonian, $P_P(s) = \exp (-s)$, while the level statistics 
of a chaotic system is given by the Wigner-Dyson distribution, 
$P_{WD}(s)=(\pi s/2) \exp (-\pi s^2 /4)$ \cite{weide}.
In the absence of defects, that is, when we have an ideal chain,
the system is integrable and it can be 
solved with the Bethe ansatz \cite{bethe}. Its level distribution is therefore Poissonian. As random on-site magnetic fields are turned on and their mean-square amplitude starts increasing the system undergoes a transition and becomes chaotic. In this scenario we obtain a Wigner-Dyson distribution for the level spacings. 
By further increasing the mean-square amplitude, localization
eventually takes place and the distribution becomes Poissonian again.

When the states of the system become localized, the excitations in the
chain are restricted to finite regions of space. Defects even 
more different in energy finally leads to strong localization,
by which we mean that each eigenvector of the system becomes 
very close to a non-interacting multi-excitation state
$|\phi _i \rangle = |\alpha _1, ..., \alpha _L \rangle$, 
where $\alpha _k =0,1$ indicates a spin down or up, respectively,
and $L$ is the total number of sites. Each excitation is then
restricted to practically just one site of the chain.
In quantum computation, these are the states we want to operate with
and they are called quantum registers
\cite{shepe_QC}. On the other hand,
the eigenfunctions of the system in the absence of defects and
especially the eigenfunctions in the chaotic regime are 
very much spread over the chain and they correspond to 
linear superpositions of several quantum registers. The system 
in this case is said to be delocalized.

The Hamiltonian that describes our system is the following,
\begin{eqnarray}
H &=& \sum_{n=1}^{L} \frac{h_n}{2} \sigma _{n}^{z}+
\sum _{n=1}^{L-1} \frac{J_Z}{4} \sigma_n^z \sigma_{n+1}^z +
\nonumber \\
  &+& \sum _{n=1}^{L-1} \frac{J_{XY}}{4} (\sigma_{n}^x \sigma_{n+1}^x +
\sigma_{n}^y \sigma_{n+1}^y) .
\label{ham}
\end{eqnarray}
We consider only nearest neighbor interaction, $\hbar = 1$, and $\sigma^x$, $\sigma^y$, and $\sigma^z$ are Pauli matrices. Each site $n$ is subjected to a magnetic field in the 
$z$ direction, giving the energy splitting $h_{n}$. There are $L$ sites
and we deal with an open chain, which is, in our opinion, 
a more realistic model for quantum computers. The system is isotropic 
(anisotropic) when 
the coupling constant $J_Z$ for the diagonal Ising interaction 
$\sigma_n^z\sigma_{n+1}^z$ is equal (different) 
to the coupling constant $J_{XY}$
for the $XY$-type interaction 
$\sigma_n^x\sigma_{n+1}^x + \sigma_n^y\sigma_{n+1}^y$. 
This last term is responsible for delocalizing the system, 
because it propagates the excitations through the chain. It is usually 
referred to as the hopping term.

To analyze how localization affects the entanglement of a pair of qubits, we compare the number of principal components for the whole system with the concurrence for a chosen pair. The definitions for these two quantities are presented in the next section.
There we restrict ourselves to two qubits and show a brief analytical 
discussion. As expected, entanglement tends to disappear in a 
strongly localized system.
On the other hand, it becomes maximal if the level spacing of the two qubits 
is the same and it is also very large when the interaction between them 
is much larger than the difference between the two level spacings. 

In section III, we show our numerical results for a chain
with several qubits and several excitations. First, we discuss the case 
where two 
qubits have the same level spacings, which are in turn very different from 
the other qubits. If the Ising interaction 
does not exist and only the hopping term is present,
we have a situation very similar to the case
of only one excitation discussed in section II and maximal entanglement
can be obtained.
However, if the Ising term is present,
the entanglement for nearest neighbor
qubits becomes larger than for the other pairs, indicating that many-body 
effects can strongly affect entanglement between distant neighbors.

In section III, we also consider qubits with random level spacings,
which allows us to study the behavior of entanglement with chaos
and the behavior of entanglement with 
localization when several interacting excitations are present.
In the integrable but delocalized region
of an ideal chain, entanglement is large. It then decreases when the 
system becomes chaotic and even more delocalized. However, in the 
transition region between chaos and strong localization, 
where the system is actually localizing, entanglement
increases again. It is only when the system gets close
to strong localization that entanglement decreases.  
The relation between localization and entanglement is 
not so simple as one might have expected. 
But it is clear from our analysis that chaos
diminishes entanglement. We should add, however, 
that for Hamiltonian (\ref{ham})
nearest neighbor interacting qubits are able to keep their concurrence 
reasonably large even in a chaotic region. 

Conclu\-ding remarks are presented in section IV.

\section{Analytical discussion: two qubits}

In order to study quantitatively the entanglement between two pairs of qubits we should calculate their entanglement of formation ($E_{F}$). Given the density matrix $\rho_{12}$ that describes our pair of qubits, $E_{F}$ is the average entanglement of the pure states of the decomposition of $\rho_{12}$, minimized over all possible decompositions:
\begin{equation}
E_{F}(\rho_{12}) = \text{min}\sum_{i}p_{i}E(\psi_{i}),
\end{equation}
where $\sum_{i}p_{i} = 1$, $0 < p_{i} \leq 1$, and $\rho_{12} = \sum_{i}p_{i}\left| \psi_{i}\right>\left< \psi_{i}\right|$. Here $E(\psi)$ is the von Neumann entropy of either of the two qubits \cite{bennett}. Wootters \textit{et al} \cite{wootters} have shown that, for a pair of qubits, $E_{F}$ is a monotonically increasing function of the concurrence, which one can prove to be an entanglement monotone. Since the concurrence is mathematically simpler to deal with than $E_{F}$, we concentrate our efforts calculating the concurrence to study the amount of entanglement between two qubits. The concurrence between them is \cite{wootters}:
\begin{equation}
C_{12} = \text{max} 
\{ \lambda_{1} - \lambda_{2} - \lambda_{3} - \lambda_{4}, 0 \},
\end{equation}
where $\lambda_{1},  \lambda_{2}, \lambda_{3}$, and $\lambda_{4}$
are the square roots of the eigenvalues, in decreasing order, of the matrix $R =\rho_{12}\tilde{\rho}_{12}$. Here $\tilde{\rho}_{12}$ is the time reversed matrix given by
\begin{equation}
\tilde{\rho}_{12} = \left(\sigma_{y} \otimes \sigma_{y}\right) \rho^{*}_{12} \left(\sigma_{y} \otimes \sigma_{y}\right).
\end{equation} 
The symbol $\rho^{*}$ means complex conjugation of the matrix $\rho$ in the basis $\left\{ \left| 11 \right>, \left| 10 \right>, \left| 01 \right>, \left| 00 \right> \right\}$. 

The above procedure to calculate the concurrence of two qubits is used in the next section, where we deal with several qubits. There, we
trace over the qubits we are not interested in and study the reduced density matrix of the two chosen ones. For a pure bipartite system, which is the case for the eigenvectors of a two spin chain, the concurrence is simply given by:
\begin{equation}
C_{12} = 2 \left| ad - bc \right|, 
\label{simples}
\end{equation}
where $\left|\psi\right>_{12} = a \left| 11 \right> + b \left| 10 \right> + c \left| 01 \right> + d \left| 00 \right>$ is the bipartite pure state. We see that when the concurrence is zero the two qubits are not entangled and when it is $1$ they are maximally entangled (EPR-Bell states).

For an isotropic spin chain with two sites 
the Hamiltonian (\ref{ham}) can be written as:
\begin{equation}
H = \frac{1}{2}\left( h_{1} \sigma _{1}^{z} + h_{2} \sigma_{2}^{z} \right) +
\frac{J}{4} \left(\sigma_{1}^{z} \sigma_{2}^{z} + \sigma_{1}^{y} \sigma_{2}^{y} + \sigma_{1}^x \sigma_{2}^{x} \right).
\label{ham12}
\end{equation}
One can easily verify that $[H, \sigma_{1}^{z} + \sigma_{2}^{z}] = 0$, i. e., the total angular momentum in the $z$-direction is conserved. This implies that states with different number of excitations do not couple. This is evident by looking at the matrix form of $H$:
\begin{equation}
  H  =  
    \left(
	\begin{array}{cccc}
            \frac{\Sigma}{2} + \frac{J}{4} & 0 & 0 & 0 \\
	    0 & \frac{\Delta}{2} - \frac{J}{4} & \frac{J}{2} & 0 \\
            0 & \frac{J}{2} & -\frac{\Delta}{2} - \frac{J}{4} & 0 \\
	    0 & 0 & 0 & -\frac{\Sigma}{2} + \frac{J}{4}   	
	\end{array}
    \right),
\end{equation}
where $\Sigma = h_{1} + h_{2}$ and $\Delta = h_{1} - h_{2}$. The above Hamiltonian is block diagonal, which means that $\left| 11 \right>$ and $\left| 00 \right>$ are non-entangled eigenvectors of $H$. As we are interested in entanglement, we restrict our analysis to the other two eigenvectors. These two eigenvectors are obtained via the diagonalization of the above $2 \times 2$ block matrix. After a straightforward calculation we obtain the two remaining  eigenvectors and eigenvalues:
\begin{eqnarray}
\left| E_{\pm} \right> & = & \frac{1}{\sqrt{2\left(J^2+\Delta^2\right) \mp 2\Delta\sqrt{J^2+\Delta^2}}} \nonumber \\ \label{eigenvec}
 & & \times \left[ J \left|10 \right> - \left(\Delta \mp \sqrt{J^{2}+\Delta^{2}}\right) \left| 01 \right> \right],
\end{eqnarray}
\begin{equation}
E_{\pm} = -\frac{J}{4} \pm \frac{\sqrt{J^2+\Delta^2}}{2}.
\end{equation}

Using Eq.~(\ref{simples}) we get the following expression for the concurrence of the eigenvectors $\left| E_{\pm} \right>$:
\begin{equation}
C_{\pm} = \frac{1}{\sqrt{1 + \Delta^2/J^2}}. 
\label{conc12}
\end{equation}
Eq.~(\ref{conc12}) clearly shows that $C_{\pm}$ is a decreasing function of 
$\Delta$ and an increasing function of the coupling constant $J$. When 
$\Delta = 0$, that is when the two qubits have the 
same level spacing, we obtain maximal entangled states 
whether we are in the weak or strong coupling regime. For any other value 
of $\Delta$ we cannot achieve maximal entangled states, showing that the 
appearance of a defect ($h_{1}\neq h_{2})$ 
in this two-site chain reduces the amount of entanglement. 
But, even if $\Delta \neq 0$, when $J \gg \Delta$ we get 
large concurrences. This means that in the regime of strong coupling we 
can still have large entanglement.

To investigate the relationship between entanglement and localization we 
compute the number of principal components, $N_{pc}$.
This is a quantity often used to determine how much spread each eigenstate of the system is. The eigenvectors of the Hamiltonian (\ref{ham}) are written as linear superpositions of the quantum registers $\left| \phi _i \right>$, 
{\it i.e.} $\left| \psi _j \right> = 
\sum _{i=1,N} a_i^{j} \left| \phi _i \right>$, where $N$ is the total
number of eigenstates. 
The $N_{pc}$ for the eigenvector $j$ is defined as \cite{zele}
\begin{equation}
N_{pc}^j = 1/\sum_{i=1}^{N} \left| a_{i}^j \right|^{4}.
\label{npc}
\end{equation}
A delocalized system has large $N_{pc}$, while a 
strongly localized system has $N_{pc}$ very close to $1$.

Applying Eq.~(\ref{npc}) to the eigenvectors $\left| E_{\pm} \right>$ we get the following expression for $N_{pc}$: 
\begin{equation}
N_{pc}^{\pm} = \frac{1}{1-C^2_{\pm}/2}.
\label{npc2}
\end{equation}
Eq.~(\ref{npc2}) shows that $N_{pc}$ for these eigenvectors is an increasing function of the concurrence, which implies that, 
for this particular two-qubit system, the more it is delocalized the more it is entangled. When we have $C_{\pm} = 1$ (maximum entanglement), 
$N_{pc}^{\pm} = 2$. When $C_{\pm} = 0$ (no entanglement) 
we get $N_{pc}^{\pm} = 1$ (strong localization).
However, this association between delocalization and entanglement 
is not always valid and this will become clear in the 
next section, where we deal with several qubits. For the moment,
it is worthy noting that for a general superposition of the four quantum registers we may have large $N_{pc}$ but no entanglement at all. 
Even though the state 
$\left| \psi \right> = \sqrt{1/4}\left( \left| 11 \right> + 
\left| 10 \right> + \left| 01 \right> + \left| 00 \right> \right)$
is not an eigenstate of our Hamiltonian, it can still be used as a
good example of this situation, for it has $N_{pc} = 4$ but $C_{\psi} = 0$.

\section{Numerical results: several excitations}

In this section we study how
chaos and localization may affect the entanglement of a pair of qubits
when several qubits and several excitations are considered. 
Depending on the defects, our disordered system may become chaotic. To determine if a system is chaotic or not it is usual to compute its energy level spacing distribution. Notice that in the model described by Hamiltonian (\ref{ham}), the $z$ component of the total spin $\sum_{n=1}^L S_n^z$ is conserved, so states with different number of excitations are not coupled. We therefore 
analyze sectors with the same number of excitations. Since we are interested in comparing chaos and entanglement, we focus on the sector with the largest number of states, that is, the sector with $L/2$ excitations, because this is the region where chaos should set in first \cite{shepe_QC}.

In a very large system, the boundary conditions have no effects, 
but numerical calculations are limited to a finite number of sites. 
A chain with free boundaries (open chain) and 
defects only on the edges is 
known to be integrable \cite{edge}. For our numerical calculations we choose an open chain with defects of values $-J/2$ on the edges. Such values should diminish border effects. The Hamiltonian 
becomes now
\begin{equation}
\tilde{H} = H -\frac{J_Z}{2} \frac{(1 + \sigma_1 ^z)}{2}
-\frac{J_Z}{2} \frac{(1 + \sigma_L ^z)}{2}.
\label{ham2}
\end{equation}
We work with $L=12$ sites and 6 excitations, which gives us $12!/(6!\, 6!)=924$ states. 

\begin{figure}[ht]
\includegraphics[width=3.2in]{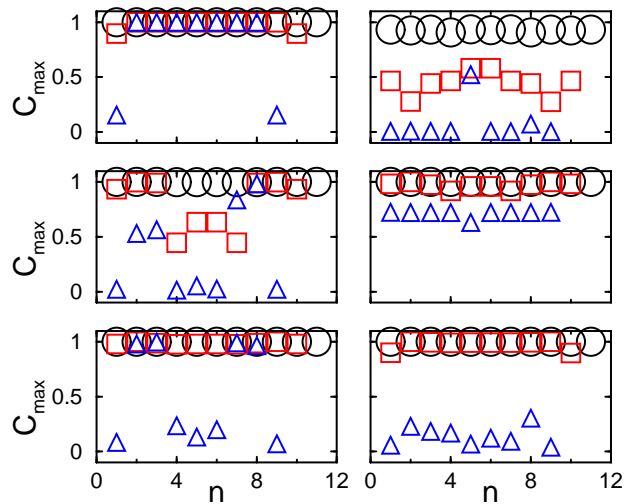}
\caption{(Color online) We show the maximum concurrence for several pairs of qubits 
with the same level spacing.
Circles indicate that the selected pairs correspond to the 
nearest neighbor qubits $(n,n+1)$, 
squares give the next-nearest neighbors $(n,n+2)$, and triangles correspond to
the next-next-nearest neighbors $(n,n+3)$. We
set $d=100$ and $J_{XY}=1$. Top left: $J_Z=0$, top right: $J_Z=J_{XY}$,
middle left: $J_Z=10 J_{XY}$, middle right: $J_Z=100 J_{XY}$,
bottom left: $J_Z=159 J_{XY}$, and bottom right: $J_Z=327 J_{XY}$.}
\label{fig1}
\end{figure}

Before analyzing how chaos can affect entanglement and
motivated by the previous section, we first check if two
qubits with the same level spacing, but now in a chain with 
 several excitations, can 
also lead to maximal entanglement.
Suppose that two selected qubits have level spacing $h+d$, 
while all the others have level spacing $h$. If we turn off the 
Ising interaction ({\it i.e.}, if $J_{Z}=0$ in Eqs.~(\ref{ham})
and (\ref{ham2})), and if $d\gg J_{XY}$, the
two selected qubits indeed lead to maximal entanglement
(except if one qubit is on the edge of the chain, in which case border
effects are noticed). This can be seen from the top left of Fig.~(\ref{fig1}),
where circles indicate that the chosen pairs are the nearest neighbor
qubits $(n,n+1)$, squares give next-nearest neighbors $(n,n+2)$,
and triangles give 
next-next-nearest neighbors $(n,n+3)$. In the figure 
we set $J_{XY}=1$ and $d=100$.
The concurrence is obtained by tracing over the qubits we are not 
interested in. We show the maximum concurrence $C_{\mbox{max}}$
obtained among
the 924 eigenstates for each selected pair. When $J_{Z}=0$,
even though we have many excitations, the situation resembles the 
case of one single excitation, 
where only the hopping term is present. Similar to this case, 
a discussion on how 
to tune chosen qubits in order to find maximally entangled states 
was done in \cite{PRA_lea}. 

However,
when the Ising interaction is on, the results can change considerably. Many-body effects now play an important role. For the 
case of an isotropic chain ($J_{Z}=J_{XY}=J$) shown on the top right
of Fig.~(\ref{fig1}), only nearest neighbors maintain
large concurrence, though not maximal. A particular case of anisotropy 
($J_{Z} > J_{XY}$), on the other hand, tends to increase the concurrence even 
for next-nearest neighbors, as can 
be noticed by comparing the graphs in the middle
of Fig.~(\ref{fig1}). On the left
$J_Z=10 J_{XY}$ and on the right $J_Z=100 J_{XY}$. 
But the anisotropy has unexpected effects on
the next-next-nearest neighbors.
In the especial case of $J_Z=d=100 J_{XY}$
shown in the middle right of Fig.~(\ref{fig1}), 
their concurrences are also increased, but 
this is not the case for large anisotropy where $J_Z\neq d$. In this
situation, the concurrence of some pairs can become very 
close to 1, while others may decrease a lot,
as can be seen in the graphs at the bottom of Fig.~(\ref{fig1}),
where arbitrary large values of $J_Z$ were chosen ($J_Z=d=159 J_{XY}$
on the left and $J_Z=d=327 J_{XY}$ on the right). Since the main 
goal of this paper, however, is to analyze the relation between chaos and
entanglement, we will let for a future publication a more 
careful study of the many-body effects due to the relative strength of the Ising ($J_Z$) and hopping ($J_{XY}$) terms in the Hamiltonian (\ref{ham}).

We now proceed to study the relation between localization and entanglement and also between chaos and entanglement in our many-body system. 
We consider an isotropic chain $(J_{Z}=J_{XY}=J)$ and set $J=1$. 
The level spacings of the qubits are obtained with random magnetic fields along the $z$ direction. They are given by $h_n = h + d_n$, where $d_n$'s are uncorrelated random numbers with a Gaussian distribution: $\langle d_n \rangle =0$ and $\langle d_n d_m \rangle =d^2 \delta _{n,m}$. 

\begin{figure}[ht]
\includegraphics[width=3.2in]{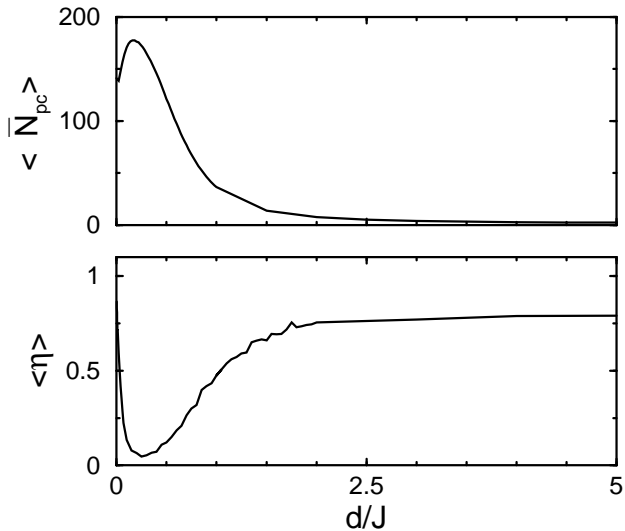}
\caption{Top: Dependence of $\langle \overline{N_{pc}}\rangle$ 
on $d/J$. The over line corresponds to an average over the 924 
eigenstates for each $d/J$ and $\langle \: \rangle$ comes from
the next average over 20 different sequences of random numbers. 
Bottom: Dependence of $\langle \eta \rangle$ on $d/J$.
Here again, $\langle \: \rangle$ indicates an 
average over 20 different sequences of 12 Gaussian random numbers.
We set $J=1$.}
\label{fig2}
\end{figure}

According to the integrability and localization of the system, 
we can identify different regions. This is shown in Fig.~(\ref{fig2}).
We calculate the average of the number of principal components
$\overline{N_{pc}}$ for the 924 eigenstates as a function of $d$. 
This is done using
a sequence of 12 Gaussian random numbers, which give the random level
spacings for the qubits. This procedure is then repeated for
20 different sequences. At the top of Fig.~(\ref{fig2}) we show 
$\langle \overline{N_{pc}}\rangle $, where $\langle \: \rangle $ 
corresponds to the average
over these 20 different sequences. We decided to work
with 20 sequences, because this is enough to
give an idea of the general behavior of the system. 
We compared the results for the concurrences of some pairs for 
more sequences and the behavior was still very similar, therefore 
justifying the use of only 20 sequences. 
At the bottom of Fig.~(\ref{fig2}) we show a quantity used to 
measure how much chaotic the system is. This parameter is defined 
as $\eta =\int _0^{s_0} 
[P(s) - P_{WD} (s)] ds /\int _0^{s_0} [P_P(s) - P_{WD} (s)] ds$, 
where $s_0 = 0.4729...$ is the intersection point of $P_P(s)$ and $P_{WD}$ 
\cite{shepe_QC,shepe97} and $P(s)$ is the energy level spacing distribution for the system under study. A regular system has $\eta =1 $ and a 
chaotic system has $\eta =0$.
Here again, $\langle \: \rangle $ indicates 
the average over 20 sequences of random numbers.

When $d=0$ the system is integrable, but delocalized. We have 
Poisson distribution $(\eta \sim 1)$, but large  
$\left< \overline{N_{pc}} \right>$. As $d$ increases the system becomes chaotic and even 
more delocalized. 
For $0<d<0.2$, we move toward
a Wigner-Dyson distribution ($\eta $ tends to 0) and 
$\left< \overline{N_{pc}} \right>$ becomes even larger.
However, as we further increase $d$ the level spacing distribution 
approaches again a Poissonian distribution and $\left< \overline{N_{pc}}\right>$ decreases.
This transition region corresponds to $0.2<d<2$.
As $d$ becomes much larger than the interaction strength $J$, the system 
becomes strongly localized, the distribution is again clearly Poissonian 
and $\left<\overline{N_{pc}}\right>$ gets very close to 1. 

\begin{figure}[ht]
\includegraphics[width=3.2in]{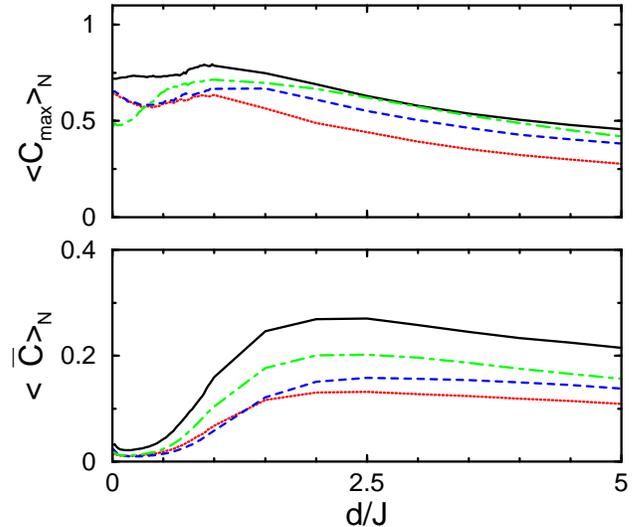}
\caption{(Color online) The top gives the dependence of $\langle C_{\mbox{max}} \rangle$ on $d/J$ and the bottom gives $\langle \overline{C} \rangle$ for nearest neighbor qubits (N). In both: solid and black line represents the pair of qubits 1-2, dotted and red line gives the pair 3-4, dashed and blue gives the pair 6-7, and dot-dashed and green represents the qubits 10-11.}
\label{fig3}
\end{figure}

\begin{figure}[ht]
\includegraphics[width=3.2in]{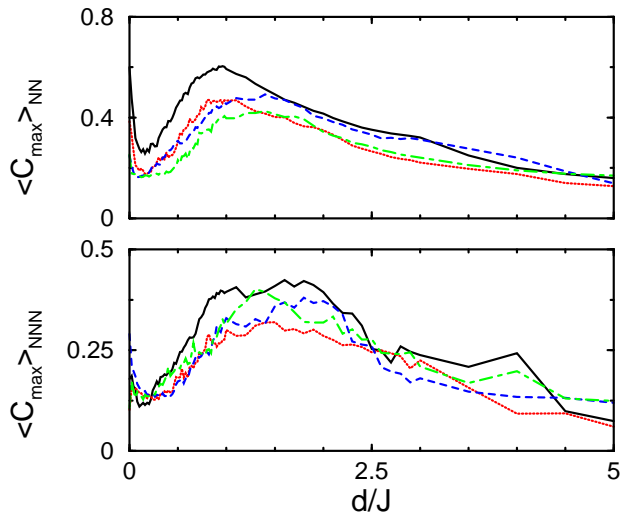}
\caption{(Color online) Top: Maximum concurrence for next-nearest neighbor qubits (NN). Solid and black line gives the pair 1-3, dotted and red is for the qubits 2-4, dashed and blue represents 3-5, and dot-dashed and green gives the pair 4-6. Bottom: Maximum concurrence for next-next-nearest neighbor qubits (NNN). Solid and black line gives the pair 2-5, dotted and red is for the qubits 4-7, dashed and blue re\-pre\-sents 5-8, and dot-dashed and green gives the pair 7-10.}
\label{fig4}
\end{figure}

We compare $\langle \overline N_{pc} \rangle $ 
and $\langle \eta \rangle $ with the concurrence for pairs 
of nearest neighbors (Fig.~(\ref{fig3})), pairs of  
next-nearest neighbors (top of Fig.~(\ref{fig4})), and 
pairs of next-next-nearest 
neighbors (bottom of Fig.~(\ref{fig4})).
For each pair we compute the maximum concurrence $C_{\mbox{max}}$ 
among the $924$ eigenstates. This is averaged over $20$ different 
sequences of random numbers. We also compute the average concurrence 
$\overline{C}$ for 
each chosen pair, which is again averaged over $20$ different 
sequences of random numbers.
The average concurrences for all pairs have 
a very similar behavior. 
Because of this and also because the values for the a\-ve\-ra\-ge 
concurrences become very small for distant pairs, 
we just show here the average concurrences 
for nearest neighbor qubits (bottom of Fig.~(\ref{fig3})).

By using Figs.~(\ref{fig2}), (\ref{fig3}), and (\ref{fig4}) we can analyze what happens to entanglement in the distinct regimes of the system.

First, let us look at  the bottom of Fig.~(\ref{fig3}).  
By comparing the average concurrences $\langle \overline{C} \rangle$ for $d=0$ and $0<d<0.2$, the region where chaos and delocalization increase,  we see that the concurrences slightly decrease. This suggests that in such systems, with nearest neighbor interactions, chaos may contribute to a decrease of the average entanglement of the qubits.
In the transition zone, $0.2<d<2$, which is the region where the system becomes less chaotic and more localized, we see a rapid increase of the concurrences. This is very interesting, because despite of being in the process of localization the average concurrences of the qubits increase. Finally for $d>2$, the region where the system is strongly localized and clearly regular (non-chaotic), the average concurrences decrease, as it should. The effect of localization on the average entanglement, therefore, depends on how far we are from the chaotic region. As explained above, localization increases entanglement when the system is moving from non-integrability (chaos) to integrability, but it decreases entanglement when the system is already strongly localized. 

The top of Fig.~(\ref{fig3}) shows that in the region of chaos ($d \sim 0.2$) the maximum concurrences for nearest neighbor qubits are larger than $\left<C_{max}\right>$ for next and much larger for next-next nearest neighbors (Fig.~(\ref{fig4})). This is explained noti\-cing that for directly coupled pairs (nearest neighbors), and therefore pairs with strong interaction, the effects of chaos are not so drastic and they keep reasonably large maximum concurrences. 
But again we verify that chaos diminishes the entanglement of any pair
of qubits.
We also see that in the region where localization is becoming strong ($d>2$) we still have reasonable values for $\left<C_{max}\right>$ for the nearest neighbor qubits. This contrasts with the behavior of $\left<C_{max}\right>$ for the next and next-next nearest neighbor qubits, where we see a faster falling $\left<C_{max}\right>$.  

From the observations of the previous paragraph we conclude that the interaction between two qubits counterbalances the destruction of their entanglement caused by chaos and also by strong localization. This interpretation is reinforced if we note that the Hamiltonian considered here has only nearest neighbor interaction terms and, as it was numerically shown, nearest neighbors maintain larger entanglement if compared to next and next-next neighbors, which are shown to be more susceptible to the effects of chaos and localization.

\section{Conclusion}

We studied how entanglement is related to chaos and localization in one dimensional Heisenberg spin chains. For this purpose we used an isotropic chain with external random magnetic fields (chain with defects) to obtain a chaotic regime.

We showed that chaos is responsible for a decrease in the entanglement (concurrence) of the nearest, next and next-next-nearest neighbor qubits. However, nearest neighbor qubits are less sensitive to chaos than the next and next-next-nearest neighbors and keep relative large entanglement in the chaotic region.

The relation between entanglement and localization is rather subtle. We found that in a strongly localized system entanglement decreases. Nonetheless, in the transition region from non-integrability (chaos) to integrability we observed that an increase in localization causes an increase in entanglement.

Finally, it was also shown that when two qubits have the same external magnetic field and this field is very large compared with the fields applied to the other sites, they lead to maximally entangled states if the Ising interaction is not present. However, this behavior is affected by the relative strength of the Ising ($J_Z$) and hopping ($J_{XY}$) terms in the Hamiltonian. Large anisotropy leads to large entanglement for nearest and next-nearest neighbor qubits, but the behavior of the next-next-nearest neighbors does not follow this or any simple trend.
\begin{acknowledgments} 
L. F. Santos acknowledges support by the NSF through grant No. ITR-0085922
and is grateful to M. I. Dykman for helpful discussions about the interplay
between interaction and disorder, its effects on localization and 
the importance of localization for quantum computation. G. Rigolin thanks
FAPESP for funding this research and C. O. Escobar thanks CNPq for partial
support.
\end{acknowledgments}


\begin{thebibliography}{99}

\bibitem{epr}  A. Einstein, B. Podolsky, and N. Rosen, Phys. Rev. \textbf{47}, 777 (1935).

\bibitem{schroedinger}  E. Schr\"{o}dinger, Proc. Camb. Phil. Soc. \textbf{31}, 555 (1935).

\bibitem{bell} J. S. Bell, Physics \textbf{1}, 195 (1964).

\bibitem{shor} P. W. Shor, SIAM J. Sci. Statist. Comput. \textbf{26}, 1484 (1997). Also available in quant-ph/9508027.

\bibitem{grover}  L. K. Grover, Proceedings, 28th Annual ACM Symposium on the Theory of Computing (STOC), May 1996, pages 212-219. Also available in quant-ph/9605043.

\bibitem{superbennett} C. H. Bennett and S. J. Wiesner, Phys. Rev. Lett. \textbf{69}, 2881 (1992).  

\bibitem{telebennett}  C. H. Bennett, G. Brassard, C. Cr\'epeau, R. Jozsa, A. Peres, and W. K. Wootters, Phys. Rev. Lett. \textbf{70}, 1895 (1993).

\bibitem{zeilinger} D. Bouwmeester, J-W. Pan, K. Mattle, M. Eibl, H. Weinfurter, and A. Zeilinger, Nature \textbf{390}, 575 (1997).

\bibitem{formation}  C. H. Bennett, D. P. DiVincenzo, J. A. Smolin, and W. K. Wootters, Phys. Rev. A \textbf{54}, 3824 (1996).

\bibitem{destilation}  C. H. Bennett, G. Brassard, S. Popescu, B. Schumacher, J. A. Smolin, and W. K. Wootters, Phys. Rev. Lett. \textbf{76}, 722 (1996).  

\bibitem{relative} V. Vedral, M. B. Plenio, M. A. Rippin, and P. L. Knight, Phys. Rev. Lett. \textbf{78}, 2275 (1997); V. Vedral, M. B. Plenio, K. Jacobs, and P. L. Knight, Phys. Rev. A \textbf{56}, 4452 (1997).

\bibitem{wootters} W. K. Wootters, Phys. Rev. Lett. \textbf{80}, 2245 (1998);
S. Hill and W. K. Wootters, Phys. Rev. Lett. \textbf{78} 5022 (1997).

\bibitem{horodecki} Rigorously the concurrence is not a measure of entanglement but an entanglement monotone. See  M. Horodecki, P. Horodecki, and R. Horodecki, Phys. Rev. Lett. \textbf{84}, 2014 (2000).

\bibitem{mark}  M. I. Dykman and P. M. Platzman, Fortschr. Phys {\bf 48}, 
9 (2000); M. I. Dykman and L. F. Santos, J. Phys. A {\bf 36}, L561 (2003);
L. F. Santos and M. I. Dykman, e-print quant-ph/0303130
(to appear in Phys. Rev. B).

\bibitem{hu} H. Li, X. Wang, and B. Hu, e-print quant-ph/038116.

\bibitem{bet_shepe} S. Bettelli and D. L. Shepelyansky, 
Phys. Rev. A {\bf 67}, 054303 (2003) and references therein.

\bibitem{fazio} S. Montangero, G. Benenti, and R. Fazio, Phys. Rev. Lett \textbf{91}, 187901 (2003).

\bibitem{Miller} P. A. Miller and S. Sarkar, Phys. Rev. E \textbf{60}, 1542 (1999).

\bibitem{Bandyopadhyay} J. N. Bandyopadhyay and A. Lakshminarayan, Phys. Rev. Lett. \textbf{89}, 060402 (2002).

\bibitem{integ_lea} L. F. Santos, e-print cond-mat/0310035.

\bibitem{weide} T. Guhr, A. M\"uller-Groeling, and H.A. Weidenm\"uller,
Phys. Rep. {\bf 299}, 190 (1998).

\bibitem{bethe} H. A. Bethe, Z. Phys. {\bf 71}, 205 (1931);
C. N. Yang and C. P. Yang, Phys. Rev. {\bf 150}, 321, 327 (1966);
M. Karbach and G. M\"uller, e-print cond-mat/9809162;
F. C. Alcaraz, M. N. Barber, and M. T. Batchelor,
Ann. of Phys. {\bf 182}, 280 (1988).

\bibitem{shepe_QC} B. Georgeot and D. L. Shepelyansky, Phys. Rev. E {\bf 62},
3504 (2000).

\bibitem{bennett} C. H. Bennett, H. J. Bernstein, S. Popescu, and B. Schumacher, Phys. Rev. A \textbf{53}, 2046 (1996). 

\bibitem{zele} V. Zelevinsky, M. Horoi, and B. A. Brown,
Phys. Lett. B {\bf 350}, 141 (1995).

\bibitem{edge} F. C. Alcaraz, M. N. Barber, M. T. Batchelor, 
R. J. Baxter, and G. R. W. Quispel, J. Phys. A {\bf 20}, 6397 (1987).

\bibitem{PRA_lea} L. F. Santos, Physical Review A {\bf 67}, 062306 (2003).

\bibitem{shepe97} P. Jacquod and D. L. Shepelyansky, Phys. Rev. Lett. {\bf 79},
1837 (1997). 

\end{thebibliography}
\end{document}